\documentclass[11pt]{iopart}

\expandafter\let\csname equation*\endcsname\relax

\expandafter\let\csname endequation*\endcsname\relax
\usepackage{amsmath}
\usepackage{graphicx}
\usepackage{bm}
\usepackage{color}
\usepackage{MnSymbol}
\usepackage{enumerate}
\usepackage{bbold}
\usepackage{subfigure}
\usepackage{listings}
\usepackage{stmaryrd}
\usepackage{multirow}
\usepackage{soul}

\newcommand{\beq}{\begin{equation}}
\newcommand{\eeq}{\end{equation}}

\newcommand{\dg}{^\dagger}

\newcommand{\smallfrac}[2]{\mbox{$\frac{#1}{#2}$}}

\newcommand{\half}{\smallfrac{1}{2}}
\newcommand{\op}[1]{\hat{ #1}}                

\renewcommand{\c}{_{\text{C}}}

\newcommand{\past}[1]{\overleftarrow{#1}}
\newcommand{\fut}[1]{\overrightarrow{#1}}
\newcommand{\both}[1]{\overleftrightarrow{#1}}
\newcommand{\fil}{_{\text F}}
\newcommand{\sm}{_{\text S}}
\newcommand{\god}{_{\text T}}
\newcommand{\subx}{_{\text X}}
\newcommand{\suby}{_{\text Y}}
\newcommand{\subn}{_{\text N}}
\newcommand{\subp}{_{\Phi}}
\newcommand{\subq}{_{\text Q}}
\newcommand{\suba}{_{\text K}}
\newcommand{\subb}{_{\text M}}
\newcommand{\subc}{_{\text H}}
\newcommand{\subss}{_\text{ss}}

\newcommand{\cxx}{\textcolor[gray]{0}{dX-dX}}
\newcommand{\cyy}{\textcolor[gray]{0}{dY-dY}}
\newcommand{\cyn}{\textcolor[gray]{0}{dY-dN}}
\newcommand{\cnn}{\textcolor[gray]{0}{dN-dN}}
\newcommand{\cxy}{\textcolor[gray]{0}{\text{\st{dX-dY}}}}
\newcommand{\cxn}{\textcolor[gray]{0}{\text{\st{dX-dN}}}}
\newcommand{\cxxx}{\textcolor[gray]{0}{\text{\st{dX-dX-dX}}}}
\newcommand{\cyyy}{\textcolor[gray]{0}{dY-dY-dY}}
\newcommand{\cnnn}{\textcolor[gray]{0}{dN-dN-dN}}
\newcommand{\cyny}{\textcolor[gray]{0}{dY-dN-dY}}
\newcommand{\cnyn}{\textcolor[gray]{0}{dN-dY-dN}}
\newcommand{\cxyx}{\textcolor[gray]{0}{dX-dY-dX}}
\newcommand{\cyxy}{\textcolor[gray]{0}{\text{\st{dY-dX-dY}}}}
\newcommand{\cxnx}{\textcolor[gray]{0}{dX-dN-dX}}
\newcommand{\cnxn}{\textcolor[gray]{0}{\text{\st{dN-dX-dN}}}}

\definecolor{nblue}{rgb}{0.06,0.3,0.73}
\definecolor{nblack}{rgb}{0,0,0}
\definecolor{nred}{rgb}{0.9,0.1,0.1}
\definecolor{nmagenta}{rgb}{0.7,0.0,0.3}
\definecolor{neditcolor}{rgb}{0.3,0.3,0.9}

\newcommand{\bo}{\bm O}
\newcommand{\bu}{\bm U}
\newcommand{\dd}{{\rm d}}
\newcommand{\dt}{\dd t}
\newcommand{\R}{R}
\newcommand{\EE}{{\rm E}}

\usepackage{array}
\newcolumntype{C}[1]{>{\centering\let\newline\\\arraybackslash\hspace{0pt}}m{#1}}

\begin{document}

\title[Quantum state smoothing: Why the types of measurements matter]{Quantum state smoothing: Why the types of observed and unobserved measurements matter}

\author{Areeya Chantasri, Ivonne Guevara, Howard M. Wiseman}
\address{Centre for Quantum Computation and Communication Technology \\(Australian Research Council), \\Centre for Quantum Dynamics, \\Griffith University, Nathan, Queensland 4111, Australia}

\ead{a.chantasri@griffith.edu.au, i.guevaraprieto@griffith.edu.au, h.wiseman@griffith.edu.au}
\vspace{10pt}
\begin{indented}
\item[\today]
\end{indented}

\begin{abstract}
We investigate the estimation technique called quantum state smoothing introduced by Guevara and Wiseman [Phys.~Rev.~Lett.~{\bf 115}, 180407 (2015)], which offers a valid quantum state estimate for a partially monitored system, conditioned on the observed record both prior and posterior to an estimation time. Partial monitoring by an observer implies that there may exist records unobserved by that observer. It was shown that, given only the observed record, the observer can better estimate the underlying true quantum states, by inferring the unobserved record and using quantum state smoothing, rather than the usual quantum filtering approach. However, the improvement in estimation fidelity, originally examined for a resonantly driven qubit coupled to two vacuum baths, was also shown to vary depending on the types of detection used for the qubit's fluorescence. In this work, we analyse this variation in a systematic way for the first time. We first define smoothing power using an average purity recovery and a relative average purity recovery, of smoothing over filtering. Then, we explore the power for various combinations of fluorescence detection for both observed and unobserved channels. We next propose a method to explain the variation of the smoothing power, based on multi-time correlation strength between fluorescence detection records. The method gives a prediction of smoothing power for different combinations, which is remarkably successful in comparison with numerically simulated qubit trajectories.
\end{abstract}


%
%
%
%
%

\section{Introduction}

Theories for open quantum systems and quantum measurements~\cite{Davies1969,Belavkin1992,BookCarmichael,BookWiseman,BookJacobs} developed during the past decades are now becoming standard tools in analysing and designing experiments for quantum technologies. Decoherence from uncontrollable or undetectable noise from the environment or measurement backaction is ubiquitous in current experiments in different platforms, such as in superconducting circuits~\cite{devoret2004superconducting,devoret2013superconducting,siddiqi2016}, ion-trap experiments~\cite{CiracZoller1995,Wineland1998}, and NV-centers~\cite{Maze2008}. However, given records from measurements that can be made on such systems, there can be different estimation strategies to best extract information about the systems. Similar estimation problems have long been considered in classical systems, and theories developed in the classical regimes  have been quite fruitful in analyzing estimation problems in the quantum regimes. In particular, the classical filtering equation \cite{BookJazwinski,Strato1960,Kushner1964}, giving a probability density function (PDF) of an estimated quantity conditioned on measurement records \textit{prior} to the estimation time, has an analogue in the well-known \textit{quantum trajectory} theory \cite{Davies1969,Belavkin1992,BookCarmichael}, describing stochastically evolving quantum states.

From the classical approach to statistics, an estimated state, i.e., a PDF, of a dynamical system can be conditioned on measurement results in three different ways: conditioning on the results prior to an estimation time (filtering), posterior to the estimation time (retrofiltering), and both (smoothing). In the regime where real-time state estimation is not required, smoothing is optimal in the sense that the whole (all-time) measurement records are used in the estimation. Carrying these ideas to the quantum regime, the analogues of the classical filtered state and the retrofiltered counterpart are the quantum trajectory and retrodictive effect \cite{BookHelstrom}, respectively. For the smoothing, there have been various approaches to applying classical smoothing to quantum systems in the literature.  Smoothing was first used for hybrid classical-quantum systems in Tsang's work \cite{Tsangsmt2009}, where an unknown classical random process was estimated from monitoring a quantum system affected by the classical process. In the later work also by Tsang \cite{Tsangsmt2009-2}, smoothing for solely quantum systems was suggested in the form of an expectation value of an observable of interest with a product of a quantum filtered state and a retrodictive effect. This approach provided a smoothed estimate of an unknown weak-measurement result of such observable, and the estimate was shown to be directly related to the \textit{weak value} \cite{ABL1964,AAV1988}. This idea was further investigated for an estimation of a hidden but arbitrary-strength measurement result, via a formalism called \textit{past quantum states} \cite{Gammelmark2013,Zhang2017,Rybarczyk2015,Tan2015,Tan2016}. 


However, as pointed out in \cite{Tsangsmt2009-2,Gammelmark2013}, the above methods could not provide a smoothed estimate of a quantum state in the usual sense, i.e., an estimated state that is Hermitian, positive semi-definite, and thus satisfies the Heisenberg uncertainty principle. This deficiency was removed recently in Ref.~\cite{Ivonne2015}, which proposed \textit{quantum state smoothing}, whereby a smoothed quantum state is introduced as a convex average of all possible \textit{true} but unknown states weighted with a classical smoothed PDF. This allows one to assign a quantum state in the usual sense to a system of interest, conditioned on measurement records both prior and posterior to the estimation time. Inspired by the semiclassical ``posterior decoding'' of Ref.~\cite{Armen2009}, Guevara and Wiseman introduced quantum state smoothing in Ref.~\cite{Ivonne2015} by considering a situation in which a quantum system of interest is coupled to several baths (see also the subsequent related work of Ref.~\cite{Budini2017}). An observer, denoted by Alice, can measure only some of the baths, yielding an \textit{observed} record $\bo$. It is then postulated that there exists another party, named Bob, who not only has access to Alice's record but also can monitor the rest of the baths, yielding a hidden or \textit{unobserved} (by Alice) record $\bu$. Therefore, given both $\bo$ and $\bu$, Bob knows the true state of the system, whereas Alice can only do her best to estimate Bob's state using her partially observed record. Using this Alice-Bob protocol, we can ask questions regarding the difference between Alice's estimated states (using smoothing and filtering) and the underlying true quantum state (Bob's state) of the system, in terms of the fidelity or the purity of the estimated states.

In the original work~\cite{Ivonne2015}, a theoretical model for a resonantly driven two-level system (a qubit with a Rabi oscillation) coupled to bosonic baths, was considered as an example. Part of the fluorescence of the driven qubit was monitored by Alice with a homodyne detection, and the remaining fluorescence was detected by Bob, using a photon detector. It was shown that Alice's \textit{smoothed} states (using the quantum state smoothing) have higher purity on average than her filtered states (i.e., the conventional quantum trajectory approach \cite{wisemanmilburn1993,Plenio1998}), and that this equates to higher fidelity on average with the corresponding Bob's true states. However, it was shown that there were noticeable differences in the purity improvement, from using different phases in Alice's homodyne detection setup. This suggested that the choices of Alice's detection (and possibly also Bob's detection) could affect the purity improvement in using smoothing over filtering. In this paper, we take up this issue and further questions regarding the \textit{smoothing power}. That is, we study qualitative measures for the improvement  smoothing offers over filtering, for various Alice-Bob detection choices. The specific questions we ask are: How does the smoothing power depend on the types of measurement performed on the qubit system? Can we qualitatively predict the smoothing power for all combinations of qubit measurements, without doing stochastic simulations?


\begin{figure}
\includegraphics[width=\textwidth]{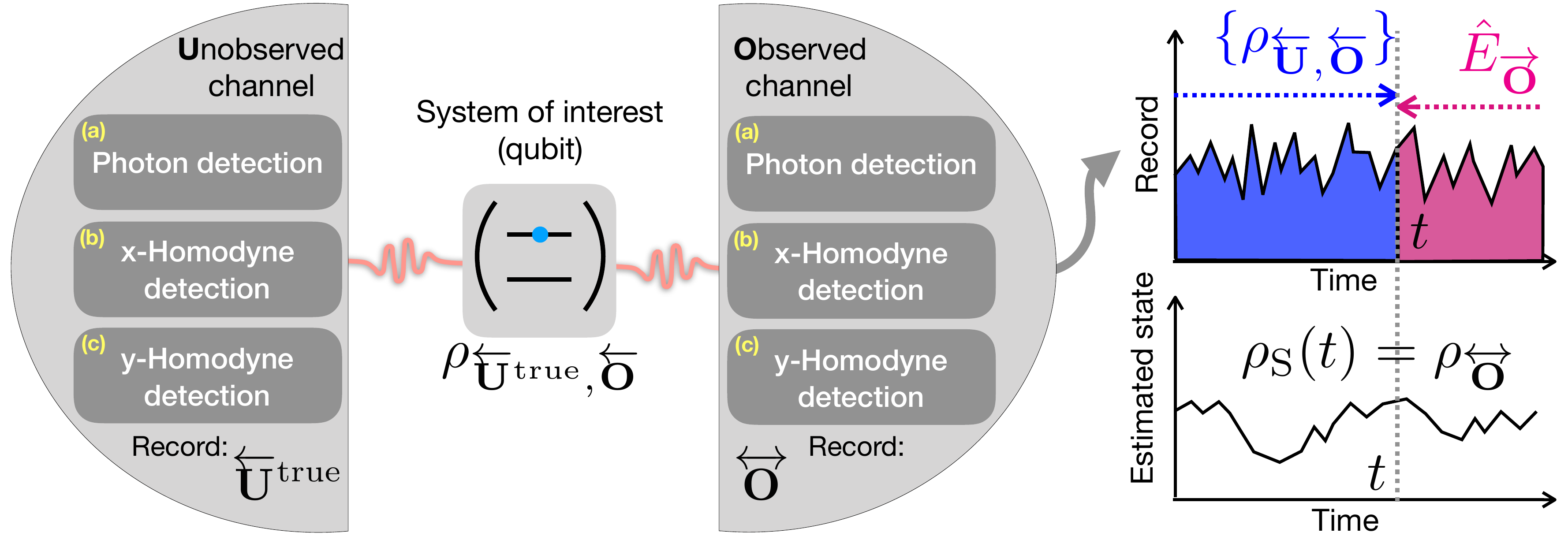}
\caption{Schematic diagrams showing the main idea of this paper. A resonantly driven two-level system (qubit) is coupled to two baths, one observed by Alice and another unobserved by her. The fluorescence detection for the two channels can be one of the three types: (a) photon detection, (b) $x$-homodyne detection, and (c) $y$-homodyne detection. Quantum state smoothing (plots on the right) is applied to the observed record, by computing a set of ``possibly true" states $\{ \rho_{\past{\bo},\past{\bu}}\}$ and the retrodictive operator ${\hat E}_{\fut{\bo}}$, and combining them according to Eq.~\eqref{eq-smtstate} to obtain a smoothed state $\rho\sm$ for Alice.}
\label{fig-intro}
\end{figure}

The specific strategy of this paper is to investigate the smoothing power for a resonantly driven qubit, considering all possible combinations of observed-unobserved measurement types for the qubit's fluorescence (see Figure~\ref{fig-intro}). The three types of fluorescence detection considered here are: (a) direct single photon detection (a jump process) and (b,c) measurement of two orthogonal quadratures using the homodyne measurement (diffusive processes). Therefore, there are nine possible combinations of the observed-unobserved detection. Each type of detection corresponds to specific backaction on the qubit system, so one would expect the smoothing power to depend on how the observed and unobserved measurements affect the state and thus each other. We quantify the smoothing power using an average purity recovery (an average of differences in purity between filtered and smoothed states) and a relative average purity recovery (a ratio between the average purity recovery and the maximum recovery possible). We then propose a predictive method for smoothing power based on \textit{correlations} between observed and unobserved records, specifically, two-time and three-time correlators, which we calculate analytically. We suggest a systematic approach to decide how much each correlator is relevant to the smoothing power. On this basis, we made our predictions for the nine observed-unobserved combinations and then test them using numerical simulations of filtered and smoothed states.

We organize the paper as the followings. We first briefly review, in Section~\ref{sec-qssreview}, the quantum trajectory theory (quantum state filtering) and quantum state smoothing theory, and introduce two measures for the smoothing power, in addition to the average purity recovery used in \cite{Ivonne2015}. In Section~\ref{sec-qtraj}, we introduce the theoretical model of a resonantly driven qubit coupled to bosonic baths, and present three possible detections for bath monitoring (unravelling) along with the stochastic master equations, which generate quantum trajectories for the filtered states. We then apply the quantum state smoothing method to this model in Section~\ref{sec-qss}, and display numerically simulated filtered and smoothed states, for a few different combinations of observed-unobserved measurements. The derivations and analyses of correlators between observed and unobserved records are presented in Section~\ref{sec-corr}, where we also make predictions on the smoothing power for all nine observed-unobserved combinations. Section~\ref{sec-numer} contains the numerical results for the purity recoveries, including errors, for all nine combinations, for comparison with the predictions made in Section~\ref{sec-corr}. The discussion and conclusion comprise Section~\ref{sec-conclude}. We also include detailed analysis for the numerical errors in Appendix A.

\section{Quantum trajectory theory and quantum state smoothing}\label{sec-qssreview}

Consider a dynamical quantum system interacting with multiple baths under the strongest Markov assumption~\cite{LiLi2018}. This allows the system's state to be monitored continuously in time via observation of the bath states. Whether the bath states are not measured, or whether the results of the measurement are ignored, the dynamics of the quantum system will be described by the same Lindblad master equation \cite{Lind1976},
\begin{align}\label{eq-lindblad}
\dd \rho(t) = -i \dt \, [{\hat H} , \rho(t)] + \dt \, \sum_{k=1}^L {\cal D}[{\hat c}_k]\rho(t),
\end{align}
where $\rho$ is the system's state matrix and ${\hat H}$ is a Hamiltonian describing any unitary dynamics of the system. The summation in the last term is over all the Lindblad superoperators ${\cal D}[\op{a}]\rho = \op{a} \rho \op{a}\dg - \half (\op{a}\dg \op{a} \rho + \rho \op{a}\dg \op{a} )$ given the Lindblad operators $\{{\hat c}_1, {\hat c}_2, ..., {\hat c}_L\}$ which describe decoherence from $L$ channels of the system-bath interaction. 

In the situation that the baths are measured continuously in time, and the results are not ignored, the system's state $\rho$ should reflect the information in the available measurement records. So far, the most commonly used approach is the quantum trajectory theory, where an estimated state of the system at any time $t$ is obtained by conditioning on the measurement records prior to that time. This is also known as quantum state filtering~\cite{Davies1969,Belavkin1992,BookCarmichael,BookWiseman}. For now, let us consider, for simplicity, a single decoherence channel $(L=1)$ and no unitary dynamics in Eq.~\eqref{eq-lindblad}, and  denote a measurement result observed from time $t$ to $t+\dt$ by $r_t$. A quantum filtered state is computed using a measurement operation via ${\tilde \rho(t+\dt)} = {\cal M}_{ r_t} \tilde{\rho}(t)$, where $\tilde{\rho}$ is the unnormalised state. Following the notation used in \cite{Ivonne2015}, we define the measurement operation ${\cal M}_{r_t}$, which acts on everything to its right, with an ostensible PDF $\wp_{\rm ost}$ such that the completeness relationship is given by $\int\!\! {\rm d}r_t \, \wp_{\rm ost}(r_t) \Tr[{\cal M}_{r_t}\rho] = 1$, for a normalised state $\rho$. Given the measurement \textit{past record}, $\past\R_{\tau} \equiv  \{ r_t : t \in [t_0, \tau) \}$, for a time of interest $\tau$ and an initial state $\rho(t_0) = \rho_0$, we get an unnormalised quantum filtered trajectory and its corresponding probability of the past record from,
\begin{align}\label{eq-filmap}
{\tilde \rho}_{\past\R_{\tau}}(\tau) = {\cal M}_{r_{\tau-\dt}} \cdots {\cal M}_{r_{t_0}} \rho_0,\\
  \wp(\past{\R}_{\tau}|\rho_0) = \Tr[{\tilde \rho}_{\past\R_\tau}(\tau)]\wp_{\rm ost}(\past{\R}_\tau),
\end{align}
where we have used $\wp_{\rm ost}(\past{\R}_\tau) = \prod_{t = t_0}^{\tau-\dt} \wp_{\rm ost}(r_t )$ as the ostensible PDF for the past record $\past\R_\tau$. By dividing a solution of Eq.~\eqref{eq-filmap} with its trace, we obtain a normalised trajectory of the quantum filtered state $\rho_{\past\R_\tau}(\tau)$. The purity of this filtered state will be unity, if the measurement results $\past\R_\tau$ are from a perfect measurement setup (no loss and no other decoherence effects) and the initial state $\rho_0$ is pure. On the other hand, if the measurement result $r_t$ is hidden or ignored, the system's state representing the lack of knowledge then decoheres according to the master equation Eq.~\eqref{eq-lindblad}, i.e., $\rho(t+\dt)  = \rho(t) + \dt {\cal D}[{\hat c}_1]\rho(t)  = \int\!\! {\rm d} r_t \, \wp_{\rm ost}(r_t) {\cal M}_{r_t}\rho(t)  = \int \!\! {\rm d}r_t \, \wp(r_t|\rho) \rho(t)$. The last expression is an average over normalised quantum states weighted by the true probabilities of the measurement results. This is why the quantum trajectory is said to generate unravellings of the Lindblad master equation. We will discuss more on different unravellings for a driven qubit example in the next section. We also note that, for the rest of the paper, we will use the word ``trajectory'' in a more general sense than the quantum filtered state trajectory, applying it to any path of a quantum state in time.


For the measurement record posterior to the estimation time, the \textit{future record}, $\fut\R_\tau \equiv  \{ r_t : t \in [\tau, T) \}$, the information can be included in the estimation as an effect operator denoted by $\op{E}$. The effect operator evolves in time according to an adjoint measurement operation ${\hat E}(t) = {\cal M}^{\dagger}_{r_t} {\hat E}(t+\dt)$. The adjoint operation is applied in a time-backward direction to an uninformative final time condition $\op{E}(T) = \op{I}$, giving a retrofiltered matrix and corresponding probability of the future record,
\begin{align} \label{eq-retfilmap}
{\hat E}_{\fut\R_\tau}(\tau) ={\cal M}^{\dagger}_{r_\tau} \cdots {\cal M}^{\dagger}_{r_{T-\dt}} \op{I},\\
 \wp(\fut{\R}_\tau|\rho_{\tau}) = \Tr[{\hat E}_{\fut\R_\tau}(\tau) \rho(\tau)]\wp_{\rm ost}(\fut{\R}_\tau),
\end{align}
where the ostensible PDF $\wp_{\rm ost}(\fut{\R}_\tau)$ is for the future record $\fut\R_\tau$ and $\rho(\tau)$ is a system's state at time $\tau$. 
It can be shown \cite{Tsangsmt2009-2} that the effect operator can be considered as a quantum analogue of the classical retrodictive likelihood function, i.e., the PDF of the future measurement record given a state at an estimation time $\tau$. 


Following Guevara-Wiseman quantum state smoothing \cite{Ivonne2015}, there are two measurement records; one is observed ($\bo$) and another is unobserved ($\bu$) by Alice. Note that we now consider a general case, where the bold letter ($\bo$ or $\bu$) denotes measurement records arising from an arbitrary number $L$ of channels in Eq.~\eqref{eq-lindblad}. Alice does not know Bob's record $\bu^{\rm true}$, but she can guess a possible $\bu$. Let us assume that an initial state of the system of interest is known. We can then construct a possible true (Bob's) state $\rho_{\past{\bo},\past{\bu}}(t)$ from the observed record $\bo$ and a possible unobserved record $\bu$ using Eq.~\eqref{eq-filmap}, with $\past\R$ replaced by $\past\bo$ and $\past\bu$. Since Alice has access only to the observed part, her \textit{filtered} state, denoted by $\rho\fil$, will then be equivalent to
\beq
\rho\fil(t)= \rho_{\past{\bo}}(t) =  {\rm E}_{\past{\bu}|\past{\bo}}[\rho_{\past{\bo},\past{\bu}}(t)] = \sum_{\past{\bu}} \wp(\past{\bu}|\past{\bo})\rho_{\past{\bo},\past{\bu}}(t),
\label{eq-filstate}
\eeq
where $\wp(\past{\bu}|\past{\bo}) \propto \wp_{\rm ost}(\past{\bu}) \Tr[ {\tilde \rho}_{\past{\bo}, \past{\bu}}(t)]$. We have used ${\rm E}_{A|B}[X]$ to represent an expected value of $X$ averaged over $A$ for a given $B$, with probability weight given by $\wp(A|B)$. We can interpret Eq.~\eqref{eq-filstate} as an average over all possible true states guessed by Alice, conditioning on her past record $\past\bo$. This also coincides with a filtered state $ \rho_{\past{\bo}}(t)$ given by Eq.~\eqref{eq-filmap}, with ${\cal M}_{o_t}$ not being purity-preserving because of the extra decoherence coming from the unobserved record $\bu$. In a similar spirit, the quantum smoothed state is defined as
\beq
\rho\sm(t) = {\rm E}_{\past{\bu}|\both{\bo}}[\rho_{\past{\bo},\past{\bu}}(t)] = \sum_{\past{\bu}} \wp(\past{\bu}|\both{\bo}) \rho_{\past{\bo},\past{\bu}}(t),
\label{eq-smtstate}
\eeq
where the only difference from the filtered state Eq.~\eqref{eq-filstate} is the weighting probability $\wp(\past{\bu}|\both{\bo}) \equiv \wp(\past{\bu}|\past{\bo},\fut{\bo}) $, which is now conditioned on both the past and future observed records. To simulate the smoothed state, we require an expression for this conditional PDF, which can be computed. This can be obtained from elementary manipulation of probability using Eqs.~\eqref{eq-filmap} and \eqref{eq-retfilmap},
\begin{align}
\wp(\past{\bu}|\both{\bo}) & \propto \wp(\fut{\bo}|\past{\bu},\past{\bo})\wp(\past{\bu},\past{\bo}), \nonumber \\
& \propto \wp_{\rm ost} (\past{\bu}) \Tr[{\hat E}_{\fut{\bo}}(t) {\tilde \rho}_{\past{\bo}, \past{\bu}}(t)],
\end{align}
where, in both lines, we have omitted any proportionality factors that are independent of the unobserved record. It is obvious that the smoothed state in Eq.~\eqref{eq-smtstate} is Hermitian and positive semi-definite, as desired. 

From comparing Eq.~\eqref{eq-filstate} and \eqref{eq-smtstate}, the smoothing equation uses information of the whole observed record $\both\bo$. Therefore, one would expect Alice's smoothed state $\rho\sm$ to estimate Bob's (true) state $\rho\god \equiv \rho_{\past{\bo}, \past{\bu}^{\rm true}}$ better than Alice's filtered state $\rho\fil$ does. To quantify the quality of conditioned estimates $\rho\c$ (which are either $\rho\fil$ or $\rho\sm$), we look at the fidelity between them and the true state, and average over all possible observed records to get a figure of merit. In a specific case when the true state is pure, it has been shown \cite{Ivonne2015} that the average fidelity is equivalent to the average purity of Alice's conditioned states, that is
\beq
\EE_{\bo}[ F[\rho\god(t),\rho\c (t)]] = \EE_{\bo}[ P[\rho\c(t)] ].
\label{eq-fidelpure}
\eeq
This is because the fidelity, as first defined by Jozsa~\cite{Jozsa1994}, reduced to $F[\rho\god, \rho\c] \equiv \Tr[\rho\god \rho\c]$ for pure state $\rho\god$, and the purity is $P[\rho\c] \equiv \Tr[\rho\c^2]$. The expected value $\EE_{\bo}[ \cdot ]$ is calculated by averaging over all realisations of observed records $\bo$. From Eq.~\eqref{eq-fidelpure}, we can then use the average purity as a measure of how well a conditioned state $\rho\c$ estimates the unknown state $\rho\god$, without the need to know what the actual $\rho\god$ is. 

For an individual observed record, the smoothed state purity at any given time can be sometimes larger, sometimes smaller, than the filtered one, resulting from fluctuation in individual noises. However, on average, we expect the smoothed purity to perform at least equal to, and generally better than, the filtered purity. In this paper, we use two measures to quantify the smoothing power, which are

\textit{1) Average Purity Recovery}
\beq
{\cal R}_{{\rm A},t} = \EE_{\bo}[ P[\rho\sm(t)]] - \EE_{\bo}[ P[\rho\fil(t)]],
\label{eq-PR}
\eeq

\textit{2) Relative Average Purity Recovery}
\beq
{\cal R}_{{\rm R},t} = \frac{\EE_{\bo}[ P[\rho\sm(t)] ] - \EE_{\bo}[ P[\rho\fil(t)]]}{\EE_{\bo}[ P[\rho\god(t)]] - \EE_{\bo}[ P[\rho\fil(t)]]}.
\label{eq-RPR}
\eeq
The first quantity describes how much purity is improved on average, from using smoothing over filtering, whereas the second quantity describes a ratio of the purity improvement over the maximum improvement possible. We will see later in the example of the driven two-level system that the bound of the average purity set by $\EE_{\bo}[ P[\rho\god(t)]]$ in \eqref{eq-RPR} can be replaced by a different value, less than one, if the dynamics of the filtered and smoothed states are confined to some subspace of quantum states. We note that the discussion in this section is general and can be applied to any type of Markovian quantum systems. But in the next sections, we turn to the applications of the smoothing technique to the afore-mentioned example. 


\section{Different unravellings for a resonantly driven qubit in vacuum bosonic baths}\label{sec-qtraj}

We follow the application of quantum state smoothing in Ref.~\cite{Ivonne2015}, considering a resonantly driven two-level system spontaneously emitting photons to coupled vacuum bosonic baths. The baths can be measured, for example, via photon detection or dyne detection \cite{Wiseman1993-2}. The Lindblad operator for the spontaneous decay of the two-level system is a lowering operator $\op{\sigma}_-$, therefore, from Eq.~\eqref{eq-lindblad} with $L=1$, we obtain a Lindblad equation of the form, 
\beq
\dd \rho = -i \dt \, [{\hat H} , \rho(t)] + \dt\, \gamma {\cal D}[\op{\sigma}_-]\rho(t),
\label{eq-mastersme}
\eeq
where we have used $\gamma$ for the total decay rate and ${\hat H} = (\Omega/2){\hat \sigma}_x$ describes resonant driving in a rotating frame, where $\Omega$ is the Rabi frequency. This Lindblad evolution can be unravelled, giving a filtered state trajectory that is dependent on the type of measurement applied on the bath and the particular measurement record obtained. In this work, we are interested in three types of bath detection (as shown in Figure~\ref{fig-intro}): (a) direct photon detection (jump record), (b) $x$-quadrature homodyne detection, and (c) $y$-quadrature homodyne detection (where (b) and (c) give diffusive records). The latter two are orthogonal quadrature measurements defined by the homodyne local oscillator phases, $\Phi = 0$ and $\Phi = \pi/2$, corresponding to measuring $\hat\sigma_x$ and $\hat\sigma_y$ observables of the qubit, respectively. These three detection schemes capture the most interesting types of backaction, which have all been seen in experiments \cite{Kimble1977,Lu1998,CamIbar2014}. In this section, for a pedagogical purpose, we consider a perfectly observed decoherence channel, where the observed measurement rate is given by $\gamma_{\rm o} = \gamma$. The discussion below follows the treatment presented in Ref.~\cite{BookWiseman}.


For photon detection, since at most one photon can be detected during an infinitesimal time $\dt$, the measurement backaction on the system's state is described by a measurement operation, ${\cal M}_{\dd J\subn}$, where $\dd J\subn = 0$ or $1$ is the measurement result, representing zero-photon or single-photon detection event, respectively. Let us define $\op{c} = \sqrt{\gamma} \op{\sigma}_-$ as a jump Lindblad operator. If a photon is detected ($\dd J\subn = 1$), the system's state collapses to its ground state through the action of the measurement operation ${\cal M}_{1} \rho = \op{c} \, \rho \,  \op{c}\dg$. For the zero-photon event, the measurement backaction is described by ${\cal M}_0 \rho = {\hat M}_0 \rho {\hat M}_0\dg$, where ${\hat M}_0 = {\hat 1} - i {\hat H} \dt - \tfrac{1}{2} \op{c}\dg \op{c} \, \dt  $, which includes the unitary evolution for the infinitesimal time $\dt$. In the time-continuum limit with a continuous record $\dd J\subn(t)$, one can write down a stochastic master equation for the photon detection,
\beq\label{eq-smejump}
\dd \rho(t) = - i \, \dt [ \hat H , \rho(t)] +\dd J\subn(t) {\cal G}[\op{c}] \rho(t) - \dt\, {\cal H}[\tfrac{1}{2} \op{c}\dg \op{c}] \rho(t),
\eeq
where, following \cite{BookWiseman}, we are using
\begin{align}
{\cal G}[\op{a}] =&\,  \frac{ \op{a} \, \rho \, \op{a}\dg}{\Tr[\op{a} \, \rho \, \op{a}\dg]} - \rho, \\
{\cal H}[\op{a}] = &\, \op{a} \rho + \rho \op{a}\dg - \Tr[\op{a} \rho + \rho \op{a}\dg ] \rho.
\end{align}

For the homodyne detection with the local oscillator phase $\Phi$, the measurement result denoted by $\dd J\subq$ during an infinitesimal time $\dt$ can take any real value. The subscript `Q' here stands for the quadrature measurement, which can be replaced by `X' or `Y' for the measurement with the phase $\Phi =0$ or $\Phi = \pi/2$, repectively. Let us define $\op{c}\subp = \sqrt{\gamma_{\rm o}} \,\op{\sigma}_- e^{-i \Phi}$ as a Lindblad operator for the diffusive measurement. The measurement backaction for this diffusive monitoring is described by an operation ${\cal M}_{\dd J\subq} \rho = {\hat M}\subq \rho {\hat M}\subq\dg$ where ${\hat M}\subq = {\hat 1} -i {\hat H} \dt - \tfrac{1}{2}  \op{c}\subp\dg \op{c}\subp \dt + \dd J\subq \, \op{c}\subp$. The stochastic master equation for a record $\dd J\subq(t)$ is given by,
\beq\label{eq-smediff}
\dd \rho(t) = - i\, \dt [ \hat H , \rho(t)] + \dt \, {\cal D}[\op{c}\subp]\rho(t) +  \dd W(t) {\cal H}[\op{c}\subp]\rho(t)
\eeq
where $\dd W$ is an infinitesimal Wiener increment related to the measurement record as $\dd J\subq(t) = \Tr[ (\op{c}\subp + \op{c}\subp\dg)\rho(t)] \dt+ \dd W(t)$. 

We note that by averaging the stochastic master equation Eq.~\eqref{eq-smejump} (or Eq.~\eqref{eq-smediff})) over all possible realisations of the measurement record $\dd J\subn$ (or $\dd J\subq$), one obtains the consistent Lindblad master equation Eq.~\eqref{eq-mastersme}.

\section{Quantum state smoothing for a resonantly driven qubit in vacuum bosonic baths}\label{sec-qss}

Given the three possible types of bath detection in Section~\ref{sec-qtraj}, we now consider the Alice-Bob smoothing protocol for the resonantly driven qubit example. Let us assume that the qubit is coupled to two baths, where Alice measures only one bath (observed channel) and Bob monitors the other bath, which is hidden from Alice (unobserved channel). Since we here are interested only in how the type of bath observation affects the smoothing power, we choose the amount of information available on both channels (observed and unobserved) to be equal. That is, we consider a qubit-bath system described by the Lindblad equation
\beq
\dd \rho(t) = {\cal L} \rho(t) \equiv - i \dt [ \tfrac{\Omega}{2} \,\op{\sigma}_x , \rho(t)] +  \dt\, \gamma_{\rm o} {\cal D}[\op{\sigma}_-]\rho(t) + \dt\,\gamma_{\rm u}{\cal D}[\op{\sigma}_-]\rho(t),
\label{eq-master}
\eeq
where the single decoherence channel in Eq.~\eqref{eq-mastersme} is now divided into two channels, one with an observed measurement rate $\gamma_{\rm o}$ and another with an unobserved measurement rate $\gamma_{\rm u}$. For simplicity, we choose $\gamma_{\rm o} = \gamma_{\rm u} = \gamma/2$. Thus, now $\op{c} = \sqrt{\gamma/2}\,\op{\sigma}_-$ and $\op{c}_{\Phi} = \sqrt{\gamma/2}\,\op{\sigma}_- e^{-i \Phi}$.

\begin{figure}
\includegraphics[width=\textwidth]{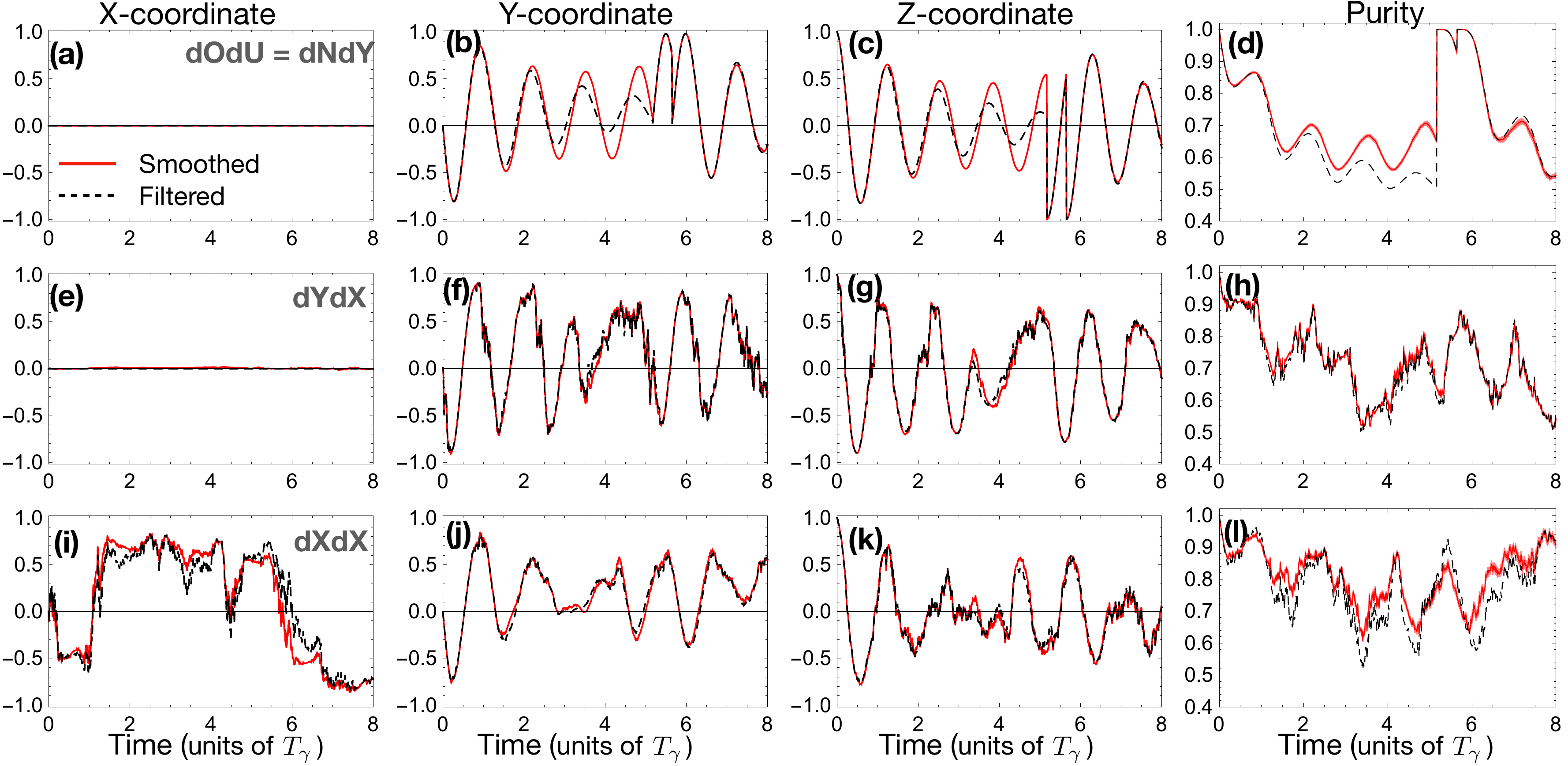}
\caption{Three sample trajectories (filtered and smoothed) shown by their Bloch sphere coordinates ($x$,$\,y$,$\,z$) and their purities. Panels (a)-(d) show filtered (black dashed) and smoothed (solid red) trajectories, where the jump record is observed and the $y$-homodyne record is unobserved, i.e., dOdU = dNdY. Panels (e)-(h) are for dOdU = dYdX, and Panels (i)-(l) are for dOdU = dXdX. The trajectories are plotted as functions of time in the unit of the decay time defined as $T_\gamma = 1/\gamma$. The Rabi oscillation frequency around the $x$-axis is $\Omega = 5 \gamma$. The measurement rate for both dO and dU channels is $\gamma/2$, the total time is $T = 8  \,T_\gamma$, and the initial state for all cases is the qubit's excited state. To obtain each smoothed quantum state, we use $10^4$ realisations of unobserved records generated randomly with ostensible statistics~\cite{Ivonne2015,Ivonne2019}. The errors from using a finite-size ensemble of unobserved records are shown only for the purity plots (see \ref{sec-appcov}), in panels (d), (h), and (l).}
\label{fig-indtraj}
\end{figure}


Now, let us define ``dO'' and ``dU'' as shorthand for observed and unobserved records, which can be one of the three options: 
\begin{equation}
\begin{split}
{\rm dN}& \Rightarrow  \text{ jump records, $\dd J\subn$}, \\
{\rm dX} &\Rightarrow \text{ $x$-homodyne ($\Phi = 0$) records, $\dd J\subx$}, \\
{\rm dY} & \Rightarrow \text{ $y$-homodyne ($\Phi = \pi/2$) records, $\dd J\suby$}.
\end{split}
\label{eq-def}
\end{equation}
Therefore, in total, there are nine combinations, denoted by ``dOdU'' (observed-unobserved), which are dNdN, dNdX, dNdY, dXdN, dXdX, dXdY, dYdN, dYdX, dYdY. 

We first examine the filtered and smoothed trajectories of the qubit subject to a few different pairs of observed and unobserved measurements via numerical simulation. We used the simulation technique of Ref.~\cite{Ivonne2015}, detailed in Ref.~\cite{Ivonne2019}. We show examples of individual qubit trajectories in Figure~\ref{fig-indtraj} by their three Bloch sphere coordinates and their corresponding purities. We chose three sample pairs of dOdU (observed-unobserved): dNdY, dYdX and dXdX.


In the first example, dNdY [the first row, panels (a)-(d)], Alice observes only jump records; therefore, her filtered state evolution (dashed black) exhibits damped oscillations with discontinuous jumps at certain times (in this case, jumps happen at $t \approx 5.14 \, T_\gamma$ and $t \approx 5.60 \, T_\gamma$, where $T_\gamma = 1/\gamma$). The curves are described by non-unitary dynamics, as they contain effects from no-jump backaction and dephasing from the unobserved part. The smoothed trajectory (solid red) diverges noticeably from the filtered one as time passes, though it resets to the same (ground) state after each jump. The purities in the panel (d) show the improvement from smoothing over filtering. The improvement tends to grow with time from the initial state or the jumps, because the filtered and smoothed states are equal at those points. The states are also equal at the final times, as there is no future record there to distinguish smoothing from filtering. It is worth noting that the smoothed trajectory, even though is a result of an average over $10^4$ possible true state trajectories, has almost no trace of the unobserved diffusive record.

In the second and third rows, panels (e)-(l) of Figure~\ref{fig-indtraj}, we present different scenarios involving the diffusive records. The panels (e)-(h) are for the case of dYdX, where the filtered and smoothed trajectories show fluctuations from an observed diffusive record dY. Theoretically, both the filtered and smoothed states should be confined to the $y$-$z$ plane of the Bloch sphere, because the initial state, the Rabi oscillation, the backaction from the observed $y$-homodyne record, and the decay from the unobserved evolution do not break the symmetry around the $x$-axis. However, since the unobserved record dX can bring the state outside of the $y$-$z$ plane, one can see tiny non-zero values in the $x$-coordinate of the smoothed state in the panel (e), resulting from finite size of ensemble of unobserved record used in averaging the possible true states in Eq.~\eqref{eq-smtstate}. In the last row, panels (i)-(l), an explicit non-zero component in $x$-coordinate is presented as a result of observing a dX record. 

From the three row of examples, one can see that the improvement in the purity of the smoothed states from the filtered states are noticeably different in the different cases. Note that in the latter two cases (dYdX and dXdX), the smoothed states are not always more pure than their corresponding filtered states. This phenomena was previously found in Ref.~\cite{Ivonne2015} for other cases (which are dYdN and dXdN) where diffusion is observed.

\begin{figure}
\includegraphics[width=\textwidth]{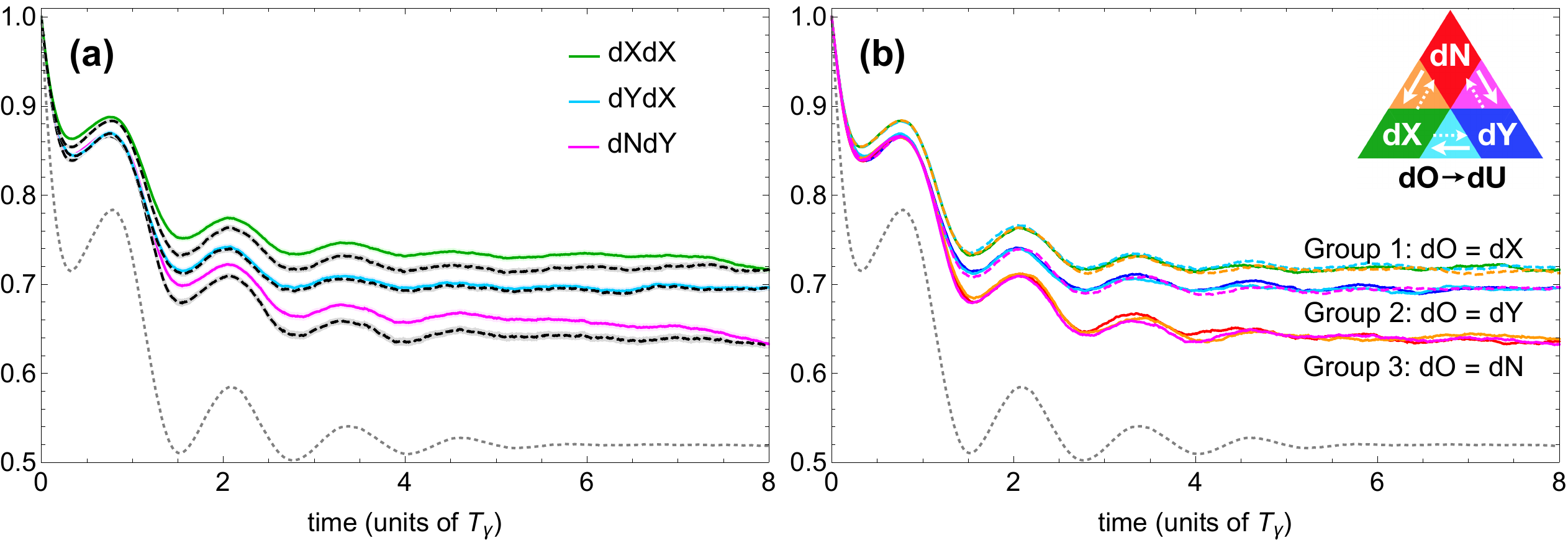}
\caption{Average purity of filtered and smoothed trajectories from 3000 realisations of observed records. (a) Average purity of smoothed trajectories (solid coloured lines) compared with their filtered counterpart (adjacent dashed black lines) for three dOdU combinations: the top pair (dXdX), the middle pair (dYdX), and the bottom pair (dNdY), where associated error bars are shown in coloured bands behind these curves. (b) Average purity of \textit{filtered} trajectories for all nine dOdU combinations, which can be categorized into three groups based on dO alone, as expected. The colour legend is read according to the labels and the arrow, for example, dOdU = dXdX (green), dOdU =  dYdN (dashed magenta), and dOdU = dNdX (solid orange). For referencing, the purity of the Lindblad dynamics solution \eqref{eq-master} is shown in both panels (dotted gray curves). The system parameters used for these plots are the same as in Figure~\ref{fig-indtraj}. Time is shown in units of the decay time $T_\gamma = 1/\gamma$.} 
\label{fig-averagetraj}
\end{figure}

We then consider average dynamics from 3000 sets of observed-unobserved records and compute average purities of smoothed and filtered states, i.e., $\EE_{\bo}[ P[\rho\sm]]$ and $\EE_{\bo}[ P[\rho\fil]]$. The average purities, for the three example cases (dNdY, dYdX and dXdX) are presented in Figure~\ref{fig-averagetraj}(a), where we also compare them to the purity of the Lindblad dynamics solution. For two cases (dNdY and dXdX), we can now see a clear improvement in the average purity of the smoothed states over the filtered states at all time (except the end points), in contrast to the result for individual trajectories in Figure~\ref{fig-indtraj}. For the third case, dYdX, a very slight improvement  is visible, but it is within the error bars of the simulations. One can also say with confidence that the dNdY combination shows the largest purity recovery of the three. 

In Figure~\ref{fig-averagetraj}(b), we show the average purity of filtered states of all nine dOdU combinations, for two reasons. First, it is to verify that the type of unobserved measurements does not affect the purity of the filtered states. This can be seen from the plot in that the averages can be categorized into three groups (label group 1,2,3) according to the type of the observed record. These coincide, within error bars (of size comparable to those shown in Figure~\ref{fig-averagetraj}(a), with each group  being distinct. Second, it is to serve as a reference guide when one considers the relative purity recovery Eq.~\eqref{eq-RPR}, as one of the two measures of the smoothing power. The average purity for the filtered state has its highest value when observing dX, second highest when observing dY, and lowest when observing dN. 

\section{Correlations between measurement records as predictors for smoothing power}\label{sec-corr}

From the previous section, we have seen that the improvement in the average purity of the smoothed state in comparison to that of the filtered state is different for the three examples dNdY, dYdX, and dXdX. However, before we delve into all the nine combinations of dOdU, we should first try to understand what leads to the purity improvement and see if we can predict the degree of recovery in a systematic way. As we learned from Section~2, the smoothed and filtered states are distinct precisely in how the probability weights for unobserved records are conditioned on observed records. Therefore, one would expect the purity recovery obtained from using smoothing over filtering to be dependent on how the two records are correlated. As an extreme scenario to build intuition, if both records were uncorrelated, i.e., all orders of correlation functions between records $\bo$ and $\bu$, were zero, one would not expect the smoothing to give any advantage over filtering at all. This is because the weighting factors $\wp(\past{\bu}|\past\bo) = \wp(\past{\bu}|\both\bo) = \wp(\past{\bu})$ in both cases are the same, independent of whether it is conditioned on the whole record or only the past record. On the other hand, if $\bo$ and $\bu$ records were highly correlated, by conditioning on the whole observed record, the smoothed state could estimate the true state state significantly better than the filtered one. This suggests that considering the correlation between the two records will be a useful predictor for the power of smoothing.


\subsection{Correlators relevant for quantum state smoothing}

To assess the correlation between observed and unobserved measurement records, the simplest quantity to analyze is the two-time correlation function. The functions which are higher order in time can be expected to be less relevant, but as we will see, the three-time correlation functions are useful as \textit{tie-breakers}.  Correlation for measurement records, by definition, can be obtained from averaging over all possible record realisations, and both two- and three-time correlation functions have been measured experimentally~\cite{FosOro2000,AbmVei2009,KocSam2011,StiKum2018}. Here we acquire analytical solutions, using the average dynamics described by the Lindblad master equation, Eq.~\eqref{eq-master}. We first consider the usual correlation functions in terms of expected values of measurement records at two and three different times, and then construct specific correlators that are useful in predicting the quality of the smoothing.

Following the method in Ref.~\cite{BookWiseman} for the derivation of autocorrelation functions, e.g., $\EE[\dd J\subn(t+\tau)\dd J\subn(t)]$ and $\EE[\dd J\subq(t+\tau)\dd J\subq(t)]$, where $\EE[\cdot]$ represents an average over all possible record realisations, we here derive the cross-correlation functions for two and three time arguments. First consider the two-time correlation between a jump record $\dd J\subn$ and a diffusive record $\dd J\subq$, which has been measured in quantum optics experiments~\cite{FosOro2000,FosSmi2002}. Recalling that the jump signal at a particular time can be either $0$ or $1$, and that $\EE[\dd J\subq(t)] = {\rm Tr}[({\hat c}_{\Phi}+ {\hat c}_{\Phi}^{\dagger} ) \rho(t) ] \dt$, we obtain (using the notation defined in Section~\ref{sec-qtraj}),
\begin{align}
\EE[\dd J\subq(t+\tau)\dd J\subn(t)]=& \EE[\dd J\subq(t+\tau) | \dd J\subn(t) = 1] \times {\rm Tr}[{\hat c}^{\dagger} {\hat c} \rho(t) ] \dt,\nonumber \\ 
=& {\rm Tr}[({\hat c}_{\Phi} + {\hat c}_{\Phi}^{\dagger}) \, e^{{\cal L} \tau} {\hat c} \rho(t) {\hat c}^{\dagger} ] \dt^2,
\end{align}
knowing that the average has no contribution from when $\dd J\subn(t) = 0$. We note that Lindblad-evolution superoperator $e^{{\cal L}\tau}$ for a duration $\tau$ acts on the product of all operators to its right. One can read the right hand side of the first line as a multiplication of an average of the diffusive signal given jumps $\dd J\subn(t) = 1$ at time $t$ and the likelihood of getting such jumps. The latter is equal to ${\rm Tr}[{\hat c}^{\dagger} {\hat c} \rho(t) ] \dt$. For an opposite ordering, where diffusive records are at time $t$ and jump records are at time $t+\tau$, the two-time correlator is different but can be obtained in a similar manner \cite{BookWiseman},
\begin{align}
\EE[\dd J\subn(t+\tau) \dd J\subq(t)] = & \EE[\dd J\subn(t+\tau)]{\rm Tr}[ ({\hat c}_{\Phi}+ {\hat c}_{\Phi}^{\dagger} ) \rho(t)]\dt + \EE[\dd J\subn(t+\tau) \dd W(t) ], \nonumber \\
=& {\rm Tr}[{\hat c}^{\dagger} {\hat c} \, e^{{\cal L}\tau}({\hat c}_{\Phi} \rho(t) + \rho(t) {\hat c}_{\Phi}^{\dagger} )]\dt^2.
\end{align}
Combining the above solutions with the autocorrelators in \cite{BookWiseman}, we can generalize the form of two- and three-time correlators for any types of measurement records denoted by $\dd J\suba$, $\dd J\subb$, and $\dd J\subc$. These dummy record variables can be any of the three types defined in Eq.~\eqref{eq-def}, and we use dK, dM, dH as shorthand for the record types, which can be any of the three: dN, dX, and dY. The general forms of correlators are given by 
\begin{align}\label{eq-corrgen1}
\EE[\dd J\subb(t_2)\dd J\suba(t_1)]\subss  &= {\rm Tr}[ {\cal K}\subb e^{{\cal L}(t_2-t_1)} {\cal K}\suba\rho\subss ]\dt^2,\\
\EE[\dd J\subc(t_3)\dd J\subb(t_2)\dd J\suba(t_1)]\subss  &= {\rm Tr}[{\cal K}\subc e^{{\cal L}(t_3-t_2)} {\cal K}\subb e^{{\cal L}(t_2-t_1)} {\cal K}\suba\dt^3,
\label{eq-corrgen2}
\end{align}
where we introduced the superoperators:  ${\cal K}\subn\rho=  {\hat c} \,\rho \,{\hat c}^{\dagger} $ (for jump records dN), and ${\cal K}\subq\rho = {\hat c}_\Phi \rho +  \rho\, {\hat c}_\Phi^{\dagger}$ (for diffusive records, e.g., dX or dY). We have also replaced $\rho(t)$ with $\rho\subss $, i.e., a solution of ${\cal L}\rho\subss  = 0$, since we are most interested in the steady-state (with the subscript `ss') behavior of the system. 

The correlators or expected values presented above, however, are not yet suitable as a measure for correlation between different pairs of observed-unobserved records in the smoothing problem. This is because the units of the jump and diffusive records are not the same, and the correlators still contain contribution from averages not relevant to smoothing. To define more meaningful formulas for correlation, we need to renormalise Eqs.~\eqref{eq-corrgen1}-\eqref{eq-corrgen2} and subtract any contributions that should not be relevant in predicting  the smoothing power. For the two-time correlation, it is straightforward to identify that an irrelevant quantity is the product of individual average records. Therefore, a normalised correlator that determines purely correlation between any two records  ($\dd J\suba$ and $ \dd J\subb$, say) at two different times in the steady-state regime is,
\begin{align}
{\cal C}_2[\dd J\suba, \dd J\subb](\tau) &= \frac{\EE[\dd J\suba(t+\tau)\dd J\subb(t)]\subss  - \EE[\dd J\suba(t+\tau)]\subss \EE[\dd J\subb(t)]\subss }{{\cal N}\suba {\cal N}\subb},
\label{eq-corrtwo}
\end{align}
which is a function of time difference $\tau$ and is symmetric under the interchange of the records, i.e., ${\cal C}_2[\dd J\suba,\dd J\subb] = {\cal C}_2[\dd J\subb,\dd J\suba]$. The normalisation factor ${\cal N}$ is defined as the \textit{root mean square} signal multiplied by a factor  $\dt^{3/2}$, giving
\begin{align}
{\cal N}\subn \,&= \sqrt{{\rm E}\left[ \left( \dd J\subn/\dt \right)^2 \right] \dt^3} \,= \dt\sqrt{{\rm Tr}[{\hat c}\dg {\hat c}\rho\subss ]}, \\
{\cal N}\subq &= \sqrt{{\rm Tr}\left[ (\dd J\subq/\dt)^2 \right] \dt^3} \, = \dt \sqrt{{\rm Tr}[ ( {\hat c}_{\Phi}+ {\hat c}_{\Phi}^{\dagger} )\rho\subss ]^2 \dt + {\rm E}[ \dd W^2 ] /\dt}  \,\approx \dt,
\end{align}
for the jump and diffusive records respectively. Using these normalised factors, the correlators are now independent of the units of the measured photocurrents, as well as of the infinitesimal time $\dt$. 

\begin{figure}
\includegraphics[width=\textwidth]{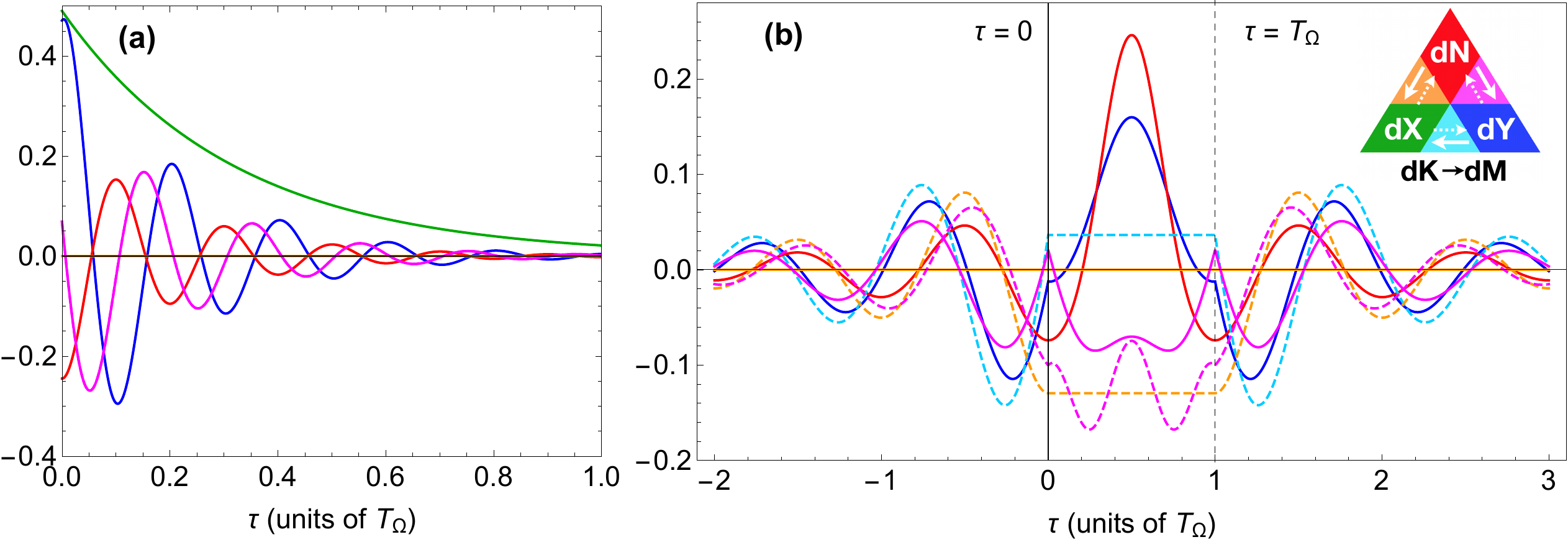}
\caption{Two-time and three-time correlators in the steady-state regime that can be used to determine the smoothing power. (a) Two-time correlators ${\cal C}_2[\dd J_{\rm K},\dd J_{\rm M}] = {\cal C}_2[\dd J_{\rm M},\dd J_{\rm K}]$ in Eq.~\eqref{eq-corrtwo} are shown as functions of $\tau$. The non-vanishing correlators are for the following: ${\cal C}_2[\dd J\subx,\dd J\subx]$ (green), ${\cal C}_2[\dd J\suby,\dd J\suby]$ (blue), ${\cal C}_2[\dd J\subn,\dd J\subn]$ (red), and ${\cal C}_2[\dd J\subn,\dd J\suby]$ (magenta). (b) Three-time correlators ${\cal C}_3[\dd J_{\rm K},\dd J_{\rm M}]$ in Eq.~\eqref{eq-corrthree} are shown as functions of $\tau$, where ${\cal T} $ is chosen to be one Rabi period. The colour legend is read in the same way as in Figure~\ref{fig-averagetraj}, but with dK and dM, representing any two types of records. The values of correlators here are used for the analysis in Table~\ref{tab-correlator}. Time $\tau$ is presented in units of the Rabi period $T_\Omega = 2\pi/\Omega$.}
\label{fig-correlator}
\end{figure}

For the three-time correlator, since we are interested only in the correlation between two records, we choose two out of the three signal variables in the expected value Eq.~\eqref{eq-corrgen2} to be from the same record with fixed time arguments between them. Then, the expected value of interest is in this form: $\EE[\dd J\suba(t+{\cal T})\dd J\subb(t+\tau)\dd J\suba(t)]$. Subtracting the irrelevant correlation from the same record, we obtain the normalised three-time correlator that can capture the correlation between two records ($\dd J\suba$ and $\dd J\subb$) in the steady-state limit,
\begin{align}
{\cal C}_3[ \dd J\suba,\dd J\subb](\tau, T) =  \frac{\EE[\dd J\suba(t+{\cal T} ) \dd J\subb(t+\tau) \dd J\suba(t) ]\subss  - \EE[ \dd J\subb ]\subss  \EE[ \dd J\suba(t+{\cal T} ) \dd J\suba(t) ]\subss }{{\cal N}\suba^2 {\cal N}\subb},
\label{eq-corrthree}
\end{align}
using the steady-state correlators defined in Eq.~\eqref{eq-corrgen1}-\eqref{eq-corrgen2}. We note that the first argument, $\dd J\suba$, of the ${\cal C}_3[\cdot,\cdot]$ definition is the record that appear twice in the correlator.


Thus we now have correlators in Eqs.~\eqref{eq-corrtwo} and \eqref{eq-corrthree} defined with unit-less measurement results and can capture solely the correlation between any two records. We show in Figure~\ref{fig-correlator}(a) and \ref{fig-correlator}(b) the two-time and three-time correlators for all combinations of records $\dd J\suba$ and $\dd J\subb$, as functions of $\tau$. There are only two out of six of the two-time correlators that are zero: ${\cal C}_2[\dd J\subx,\dd J\subn]$, and ${\cal C}_2[\dd J\subx,\dd J\suby]$. For the three-time correlators, there are three out of nine, i.e., ${\cal C}_3[\dd J\subx,\dd J\subx]$, ${\cal C}_3[\dd J\suby,\dd J\subx]$, ${\cal C}_3[\dd J\subn,\dd J\subx]$, with vanishing correlation, shown in panel (b). 


\subsection{Correlators and prediction of smoothing power}

The values of correlators, as shown in Figure~\ref{fig-correlator}, vary significantly as $\tau$ and ${\cal T} $ change and it is not obvious how the variation could contribute to the purity improvement of smoothed states over filtered states in any logical ways. Therefore, we instead focus on a parameter-independent feature, which is the vanishing or non-vanishing property of the correlators. Some correlators, such as ${\cal C}_2[\dd J\subx,\dd J\subn]$ or ${\cal C}_3[\dd J\suby,\dd J\subx]$, are zero regardless of the values of ${\cal T} $ and $\tau$. Those that do not vanish identically are non-zero for almost all values of ${\cal T}$ and $\tau$.


\begin{table}[t]
\begin{tabular}{ |>{\centering}m{1.6cm}|>{\centering}m{2cm}|>{\centering}m{7cm}|c|} 
\hline
Observed & Unobserved  & Relevant correlators: & Predicted Level of \\ 
(dO) & (dU) & dO-dU, \, dO-dU-dO, \, dU-dO-dU & Smoothing Power  \\
\hline
dX & dX  & \cxx, \, \cxxx, \, \cxxx & 3\\ 
. & dY     & \cxy, \, \cxyx, \, \cyxy &2 \\
. &  dN    & \cxn, \, \cxnx, \, \cnxn \, & 2\\
\hline
dY &  dX  & \cxy, \, \cyxy, \, \cxyx \, & 1\\ 
. &  dY     & \cyy, \, \cyyy, \, \cyyy\,& 4\\
. &  dN   & \cyn, \, \cyny, \, \cnyn\, & 4\\
\hline
dN &  dX & \cxn, \, \cnxn, \, \cxnx \, & 1 \\ 
. &  dY   & \cyn, \, \cnyn, \, \cyny\, & 4\\
. &  dN    & \cnn, \, \cnnn, \, \cnnn \,  & 4\\
\hline
\end{tabular}
\caption{Prediction for the power of smoothing using correlation strength. For all nine combinations of dO and dU, we show vanishing (struck-through) and non-vanishing correlators according to the definition in the text. We treat dO-dU correlator as giving the highest contribution to the smoothing power, then dO-dU-dO, then dU-dO-dU. Four levels of smoothing power are predicted in the last column depending on how many non-vanishing correlators are available and their predicted contribution.}
\label{tab-correlator}
\end{table}

In order to predict the power of quantum state smoothing offered by different measurement unravelling combinations, we propose the following principles. Firstly, the stronger the correlation, the better the smoothing power. We quantify correlation strength for a particular combination dOdU by postulating that the largest contribution coming from its two-time correlator, then its three-time correlators. However, there are two ways of writing the three-time correlators between dO and dU (see Eq.~\eqref{eq-corrthree}), ${\cal C}_3(\dd J_{\rm O},\dd J_{\rm U})$ and ${\cal C}_3(\dd J_{\rm U},\dd J_{\rm O})$ for the observed ($\dd J_{\rm O}$) and unobserved ($\dd J_{\rm U}$) records. We postulate that the former one is more important for smoothing, because it allows us to quantify the correlation of the dU record (at any time $t$) with the dO record both before and after $t$. That is, the correlator ${\cal C}_3(\dd J_{\rm O},\dd J_{\rm U})$ can quantify the difference between using past-future dO record conditioning as opposed to only using the past record. We then list the correlators in Table~\ref{tab-correlator} according to their contribution to the power of smoothing, indicating the vanishing property by striking through the symbols. Using shorthand notations dO-dU, dO-dU-dO, and dU-dO-dU to represent the correlators ${\cal C}_2(\dd J_{\rm O},\dd J_{\rm U})$, ${\cal C}_3(\dd J_{\rm O},\dd J_{\rm U})$, and ${\cal C}_3(\dd J_{\rm U},\dd J_{\rm O})$, respectively, we have
\begin{itemize}
\item Two-time correlators:
\begin{itemize}
\item Non-vanishing: \cxx, \cyy, \cnn, \cyn
\item Vanishing:  \cxy, \cxn
\end{itemize}
\item Three-time correlators:
\begin{itemize}
\item Non-vanishing: \cyyy, \cnnn, \cyny, \cnyn, \cxyx, \\ \cxnx
\item Vanishing: \cxxx , \cyxy, \cnxn
\end{itemize}
\end{itemize}
In the last column of Table~\ref{tab-correlator}, we categorize all nine dOdU combinations into \textit{four levels} according to their relevant strength of the correlators shown. The 4th level has its correlators all non-vanishing, as we expect from this the best purity improvement from state smoothing. At the other extreme, for the 1st level, only the three-time correlator dU-dO-dU is non-vanishing. Note that if we instead treat the three-time correlators dO-dU-dO and dU-dO-dU on equal footing, the last two levels, 2 and 1, merge to the same category.




\section{Numerical investigation}\label{sec-numer}

To test the validity of the predictions made in the previous section, we analyze qubit trajectories, numerically generated, for all nine combinations of observed and unobserved record types, and compute their average purity recovery, Eq.~\eqref{eq-PR}, and relative average purity recovery, Eq.~\eqref{eq-RPR}. For each of the nine combinations, the trajectory data includes in total $3000$ sets, where each set contains one true state trajectory $\rho\god$, one filtered state trajectory $\rho\fil$, and one smoothed state trajectory $\rho\sm$. For the calculation of each smoothed trajectory, we follow Eq.~\eqref{eq-smtstate}, using $10^4$ realisations of dU records randomly generated with ostensible statistics~\cite{Ivonne2015, Ivonne2019}. More detail of the calculation and numerical techniques used in the simulation is presented in \cite{Ivonne2019}.

\begin{figure}
\includegraphics[width=\textwidth]{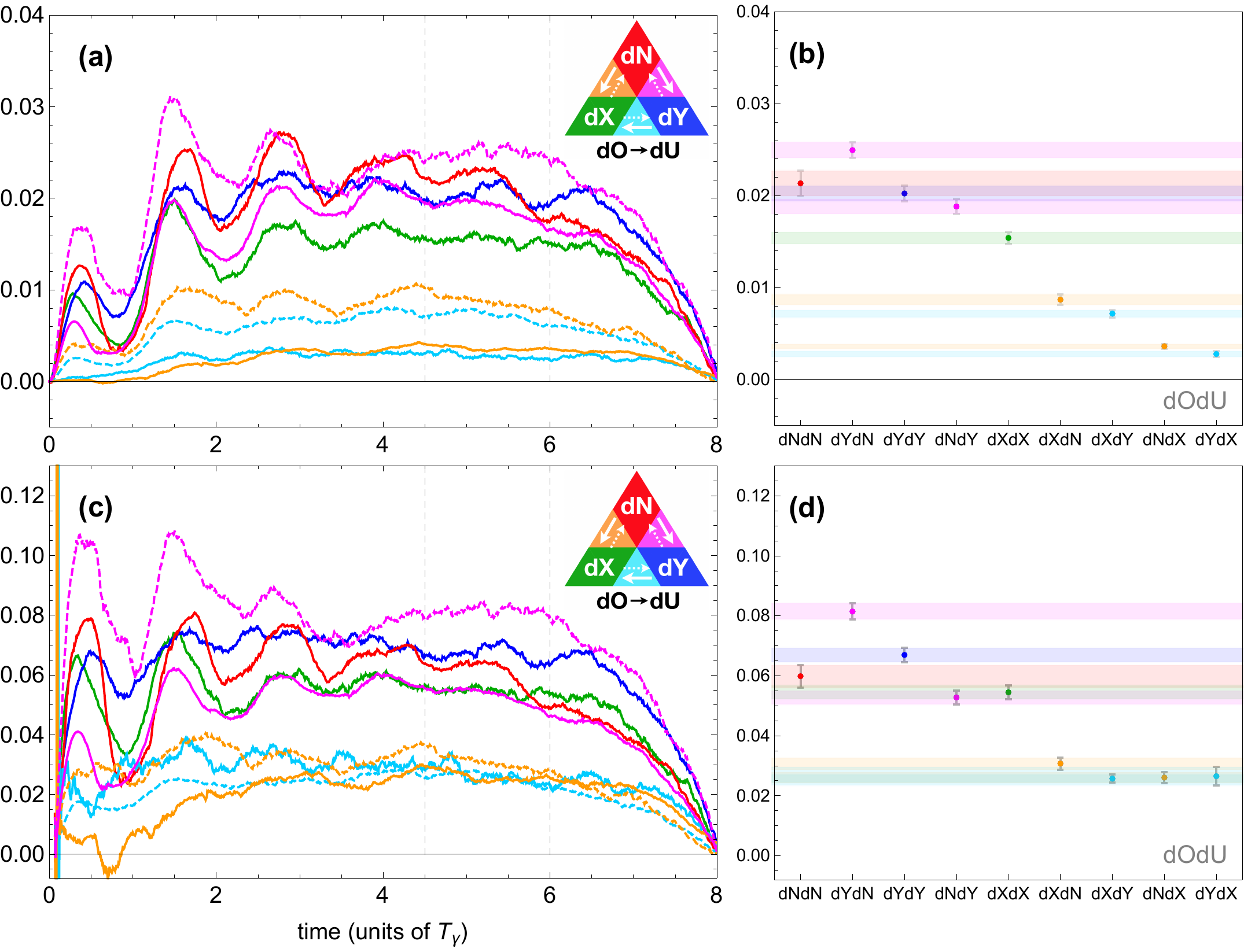}
\caption{Average purity recovery and relative average purity recovery for all nine dOdU combinations of qubit measurements. (a) Average purity recovery, Eq.~\eqref{eq-PR}, as functions of time. (c) Relative average purity recovery, Eq.~\eqref{eq-RPR}, as functions of time, where, for dNdX (solid orange) and dYdX (solid turquoise), we have substituted $P(\rho\god) \rightarrow P(\rho\god^{\rm YZ})$, i.e., the purity of the true state projection on the $y$-$z$ plane, as it is an appropriate reference for the maximum purity recovery possible for these cases. (b) and (d) are the steady-state time averages for the average purity recovery Eqs.~\eqref{eq-ss-meanPR} as in (a), and for the relative average purity recovery \eqref{eq-ss-meanRPR} as in (c), where their error bars (one standard derivation) are discussed in Appendix A.}
\label{fig-mainresult}
\end{figure}

Numerical results are presented in Figure~\ref{fig-mainresult} showing the average purity recovery in panels (a)-(b), and the relative average purity recovery in panels (c)-(d). In the panels (a) and (c), the recoveries are plotted as functions of time. Thus the values are zero at both ends, as the filtered and smoothed states are identical at $t=0$ and should be identical at the final time $t=T$ (when $\both\bo = \past\bo$  in Eq.~\eqref{eq-smtstate}). The recoveries also show transient behaviors during the time between $t_0=0$ and $t \approx 4 \,T_\gamma$, where oscillations at approximately the Rabi frequency are still visible. During the time period $\mathfrak{T}_{\rm ss} =  [4.5\, T_\gamma, 6\, T_\gamma] $, marked by vertical dashed gray lines in the plots, the results are relatively flat indicating the steady-state behavior for this period of time, before the recoveries start to converge towards zero at the final time at $T = 8 \, T_\gamma$. 

Looking at the steady-state interval defined as $\mathfrak{T}_{\rm ss}$, it is remarkable that the recovery results in Figure~\ref{fig-mainresult} can be plausibly classified into 4 groups: the top one with highest values of ${\cal R}_{{\rm A},t}$ (dNdN, dYdN, dYdY, dNdY), the second group (dXdX), the third group (dXdN, dXdY), and the last group with lowest values of the recovery (dNdX, dYdX). These groups are perfectly correlated with the prediction of levels given in Table~\ref{tab-correlator} (4th column) using the record correlators and their vanishing/non-vanishing properties. 

The average purity recovery shows only the difference in the purity of the filtered and smoothed state on average, not taking into account the value of the purity of the filtered state to begin with. In some cases, the purity recovery simply cannot be large because the filtered state is already highly pure, leaving little room for improvement by smoothing. We thus also consider the relative average purity recovery Eq.~\eqref{eq-RPR}, which is the ratio of the recovery and the difference between the filtered state's purity and the purity of the underlying true state (unit purity). However, there can be some subtleties, for cases where the true state is not a good reference for purity recovery. Examples are the two combinations dNdX and dYdX. There Alice's smoothed and filtered states are confined to the $y$-$z$ plane, but the true states can be outside of the plane, on the Bloch sphere, due to the dX record (which is unobserved by Alice). In these cases, Alice's smoothed and filtered states can never get close to the true state and therefore it makes more sense to use the true state projection on the $y$-$z$ plane as a reference in computing the relative average purity recovery. 

We show in Figure~\ref{fig-mainresult}(c), the relative recovery, using $P(\rho\god) = 1$ for all dOdU combinations except for dNdX and dYdX where we substitute $P(\rho\god) \rightarrow P(\rho\god^{\rm YZ})$ with the purity of the true state projection on the $y$-$z$ plane. Comparing to the results of the average purity recovery, the last two groups in panel (a) are no longer distinguishable in panel (c). This suggests that the reason for the dNdX and dYdX combinations have low purity recovery in (a) is because Bob was measuring dX record making his (true) state almost impossible to guess correctly by Alice, who could measure only dN or dY. Therefore, computing the purity recovery relative to the projection of Bob's state on the $y$-$z$ plane make the comparison to the dXdN and dXdY combinations more reasonable. Interestingly, the data can now be roughly categorized into two just groups (dNdN, dYdN, dYdY, dNdY, dXdX) and (dXdN, dXdY, dNdX, dYdX). This corresponds to considering only the (non-)vanishing of the two-time correlator dO-dU in Table~\ref{tab-correlator}.


The results in Figure~\ref{fig-mainresult}(a) and (c) are still quite fluctuating. We therefore consider a figure of merit from averaging the recovery during the steady-state period $\mathfrak{T}_{\rm ss}$, and its error bar. The steady-state averages for the average purity recovery and relative average purity recovery are defined as
\begin{align}
{\rm E}_{\rm ss}[{\cal R}_{{\rm A},t}] =&  \, \frac{1}{ |\mathfrak{T}_{\rm ss}|/\dt}\sum_{t \in \mathfrak{T}_{\rm ss}} \EE_{\bo}[ P[\rho\sm(t)] - P[\rho\fil(t)]], \label{eq-ss-meanPR} \\
{\rm E}_{\rm ss}[{\cal R}_{{\rm R},t}] = & \frac{1}{|\mathfrak{T}_{\rm ss}|/\dt}\sum_{t \in \mathfrak{T}_{\rm ss}} \frac{\EE_{\bo}[ P[\rho\sm(t)] ] - \EE_{\bo}[ P[\rho\fil(t)]]}{\EE_{\bo}[ P[\rho\god(t)]] - \EE_{\bo}[ P[\rho\fil(t)]]}, \label{eq-ss-meanRPR}
\end{align}
where $|\mathfrak{T}_{\rm ss}|$ is the length of the steady-state period. Since the purity of the filtered and smoothed states at each time step have their own error bars from finite-size ensemble of observed realisations (and also finite-size hypothetical unobserved realisations, for the smoothed state), we need to take all these errors into account and perform a full error analysis for the time-averaging. We present the details of the error analysis in Appendix A. The steady-state averaged recoveries with error bars in Figure~\ref{fig-mainresult}(b) and (d) have made the separate grouping much more apparent than the time-dependent plots in (a) and (c). We also found that the dYdN combination has an unexpectedly high smoothing power; it could possibly be categorised as a separate group if one had a more refined theoretical prediction. However, the prediction so far has already been quite impressive, given that it only used the information of correlation strengths between observed and unobserved measurement records.



\section{Discussion and conclusion}\label{sec-conclude}

We have investigated quantum state smoothing \cite{Ivonne2015}, particularly for a resonantly driven qubit coupled to bosonic baths, concentrating on calculating how much the smoothing helps improve the quality of state estimation over the filtering. In particular, we analyse, for the first time, how the improvement depends on the types of bath measurement, by Alice and Bob, respectively. Here we are referring to the formulation of quantum state smoothing as an Alice-Bob protocol, where the observer Alice has access to only the observed record $\bo$, whereas the hypothetical observer Bob has access to both Alice's record and the record $\bu$ unobserved by Alice. Alice is trying to estimate Bob's state, which is pure, and which can thus be considered the true state of the system.

We considered three detection schemes for the qubit's fluorescence, and therefore nine combinations of the observed-unobserved (dOdU) records: dNdN, dNdX, dNdY, dXdN, dXdX, dXdY, dYdN, dYdX, and dYdY, where dN, dX, and dY refer to photon detection records, $x$-homodyne, and $y$-homodyne detection records, respectively. Alice's smoothed quantum states have higher purity on average than her corresponding filtered quantum states, meaning that her smoothed states have higher fidelity to their corresponding Bob's (true) states, for all nine combinations. The smoothing power was then defined as a measure for the improvement, based on the amount of purity recovery the smoothing offers over the quantum filtering. Two recovery measures were analysed: the average purity recovery and the relative average purity recovery.  We found that the smoothing power is strongly dependent on the types of observed-unobserved detection. 

We developed a systematic method using correlation strength between observed and unobserved measurement records, modifying the conventional two- and three-time correlation functions, to be able to predict the smoothing power for all measurement combinations. We found that the correlation analysis correctly determined the levels of smoothing power among all nine combinations, from the prediction in Table~\ref{tab-correlator} and numerical results in Figure~\ref{fig-mainresult}. The highest smoothing power (levels 3 and 4 in Table~\ref{tab-correlator}) are any combinations of the same measurement types or the combinations involving the $y$-homodyne measurement and the photon detection. The worst combinations are ones that consist of single $x$-homodyne measurement. 

We note that even though this work is only concerned with the particular example of a qubit coupled to bosonic baths, we expect that the insights gained here will be applicable to other, quite different, physical systems, such as Linear Gaussian systems \cite{Zhang2017,Laverickarxiv}. Moreover, the ability to predict smoothing power beforehand could also be helpful in justifying the use of quantum state smoothing in estimating quantum state for experimental systems in lossy environment. There are many other interesting questions worth further investigating, such as: Would the smoothing power be maintained at the same levels if Bob's detection scheme was guessed wrongly by Alice? How does the fidelity of the smoothed state estimate compare with other state estimation methods using past-future information \cite{Chantasri2013,Weber2014}.


\ack
We acknowledge the traditional owners of the land on which this work was undertaken at Griffith University, the Yuggera people. This research is funded by the Australian Research Council Centre of Excellence Program CE170100012. AC acknowledges the support of the Griffith University Postdoctoral Fellowship scheme.

\appendix

\section{Error analysis}

We here discuss the error analysis used in computing the numerical error bars presented in Figure~\ref{fig-averagetraj} and \ref{fig-mainresult}. We discuss errors that could appear at any stages of the numerical averaging with finite-size ensembles. We start with errors coming from the calculation of an individual smoothed quantum state (involving an average over all possible unobserved records), then errors in the average purities (averaging over all possible observed records), and, finally, errors from averaging over the steady-state interval.

\subsection{Covariance of smoothed states}\label{sec-appcov}
The smoothed state is a result of a weighted average over all possible unobserved records, as in Eq.~\eqref{eq-smtstate}; therefore, using a finite-size ensemble of unobserved records (in this work, we used $N_{\bu} = 10^4$) should result in a finite error of the average. Considering a weighted average (mean) of a single random variable of the form: $\overline{x} = \sum_{i=1}^N w_i \,x_i /\sum_i w_i$, the variance of the {\em mean} (VOTM) is given by
\begin{equation}\label{eq-votm}
(\delta \overline{x})^2= \frac{\sum_i w_i^2}{(\sum_i w_i)^2}\sigma^2,
\end{equation}
where the variance of the population is
\begin{equation}
\sigma^2=\frac{\sum_i w_i x_i^2}{\sum_i{w_i}}-\overline{x}^2.
\end{equation} 
Note that if all the weights were the same, then we would arrive at a simple result $(\delta \overline{x})^2 = \sigma^2/N$.

Given the definition of the smoothed quantum state in Eq.~\eqref{eq-smtstate} and the definition of VOTM for a single variable above, we therefore can write a covariance matrix for the smoothed state as,
 \begin{equation}\label{eq:covsqs}
{\rm CoV}(\rho\sm(t)) \equiv \frac{\sum_{\past{\bf U}_{t}}\wp(\past{\bf U}|\both{\bo})^2}{\big(\sum_{\past{\bf U}_{t}}\wp(\past{\bf U}|\both{\bo})  \big)^2} \left(\sum_{\past{\bf U}_{t}} \wp(\past{\bf U}|\both{\bo})  \left(\rho_{\past{\bf O}_{t},\past{\bf U}_{t}}(t)\otimes \rho_{\past{\bf O}_{t},\past{\bf U}_{t}}(t)\right) -\rho\sm(t)\otimes\rho\sm(t)\right).
\end{equation}
We can use this matrix to calculate uncertainties for any observables, for example, the VOTM for the $y$-component of the smoothed qubit's state is given by
\begin{equation}
(\delta y\sm)^2=\Tr[(\op{\sigma}\suby\otimes\op{\sigma}\suby) \, {\rm CoV}(\rho\sm(t))].
\end{equation}
We can then calculate the variance of the smoothed state's purity defined as $P[\rho\sm] = \Tr[\rho\sm^2] = \tfrac{1}{2}(1+x\sm^2 + y\sm^2 +z\sm^2)$. From the error propagation rules, we get
\beq
(\delta P[\rho\sm(t)])^2 = x\sm^2 (\delta x\sm)^2 + y\sm^2 (\delta y\sm)^2 + z\sm^2 (\delta z\sm)^2, \label{eq-varPS}
\eeq
where we have presented this error for the smoothed state's purity in Figure~\ref{fig-indtraj}(d), (h), and (l).

\subsection{Variance of averaged purities}

The error bars presented in Figure~\ref{fig-averagetraj} are for the average purities of the filtered and smoothed states. To simplify the notations, let us use $\bar{P}\fil(t) = \EE_{\bo}[P[\rho\fil(t)]]$ and $\bar{P}\sm(t) = \EE_{\bo}[P[\rho\sm(t)]]$ for the averages of the filtered and smoothed quantum states, respectively. Using \eqref{eq-votm}, the variances of the averages are calculated from, 
\begin{align}
(\delta {\bar P}\fil(t))^2 = &\, \frac{1}{N_{\bo}}\EE_{\bo}[(P[\rho\fil(t)]- \bar{P}\fil(t)])^2],\\
(\delta {\bar P}\sm(t))^2 = & \, \frac{1}{N_{\bo}}\EE_{\bo}[(P[\rho\sm(t)]- \bar{P}\sm(t)])^2] \, +  \frac{1}{N_{\bo}}\EE_{\bo}[(\delta P[\rho\sm(t)])^2],\label{eq-varAvPS}
\end{align}
for each time $t$, with the $N_{\bo} = 3000$ ensemble of observed records. The second term of \eqref{eq-varAvPS} is a contribution from the variance of the smoothed purity \eqref{eq-varPS} for every single realisation of $\bo$ in the ensemble.

\subsection{Variance of steady-state averages}

In Figure~\ref{fig-mainresult}(b) and (d), we plotted the steady-state averages of ${\cal R}_{{\rm A},t}$ and ${\cal R}_{{\rm R},t}$ and their associated error bars. The average purity recovery and the relative average purity recovery are defined as in Eqs~\eqref{eq-ss-meanPR} and \eqref{eq-ss-meanRPR},
\begin{align}
{\cal R}_{{\rm A},t} = &\EE_{\bo}[ P[\rho\sm(t)]] - \EE_{\bo}[ P[\rho\fil(t)]] = \bar{P}\sm(t) - \bar{P}\fil(t),\\
{\cal R}_{{\rm R},t} = &\frac{\EE_{\bo}[ P[\rho\sm(t)] ] - \EE_{\bo}[ P[\rho\fil(t)]]}{\EE_{\bo}[ P[\rho\god(t)]] - \EE_{\bo}[ P[\rho\fil(t)]]} = \frac{\bar{P}\sm(t) - \bar{P}\fil(t)}{\bar{P}\god(t) - \bar{P}\fil(t)}.
\end{align}
Let us denote the steady-state averages by ${\cal R}_{\rm A,ss} = \EE\subss[{\cal R}_{{\rm A},t}]$ and ${\cal R}_{\rm R,ss}= \EE\subss[{\cal R}_{{\rm R},t}]$. The error bars presented in Figure~\ref{fig-mainresult}(b) and (d) are the square roots of the variances of the steady-state averages obtained from
\beq
(\delta {\cal R}_{\rm ss})^2= \,  \frac{1}{N_{\rm ss} |\mathfrak{T}_{\rm ss}|/\dt}\sum_{t \in \mathfrak{T}_{\rm ss}}({\cal R}_t - \EE\subss[{\cal R}_t])^2 + \frac{1}{N_{\rm ss}|\mathfrak{T}_{\rm ss}|/\dt}\sum_{t \in \mathfrak{T}_{\rm ss}} (\delta {\cal R}_t)^2,\label{eq-ss-err}
\eeq
where $\delta{\cal R}_{\rm ss} \in \{ \delta{\cal R}_{\rm A,ss}, \delta{\cal R}_{\rm R,ss} \}$, ${\cal R}_t \in \{ {\cal R}_{{\rm A},t}, {\cal R}_{{\rm R},t}\}$, and $\delta{\cal R}_t \in \{ \delta{\cal R}_{{\rm A},t}, \delta{\cal R}_{{\rm R},t}\}$. The first term of \eqref{eq-ss-err} describes the uncertainty contributed from the fluctuation of ${\cal R}_t$ within the steady-state duration, whereas the second term describes a contribution from variances of individual ${\cal R}_t$'s for $t \in \mathfrak{T}_{\rm ss}$. We also include an effective number of independent samples $N\subss$ in calculating the variance of the steady-state average, because values in consecutive timesteps are not completely independent of each other. Assuming that qubit state trajectories are uncorrelated after one correlation time $t_{\rm corr}$, the effective number of independent samples can then be approximated as $N\subss  \approx |\mathfrak{T}_{\rm ss}|/t_{\rm corr} +1$, where $|\mathfrak{T}_{\rm ss}|$ is the steady-state duration. We approximate $t_{\rm corr}$ by $T_{\gamma} = \gamma^{-1}$, the inverse of the system's decay rate. 

The variance $(\delta {\cal R}_t)^2$ in \eqref{eq-ss-err} for both ${\cal R}_{{\rm A},t}$ and ${\cal R}_{{\rm R},t}$ should be calculated carefully. The former one, the average purity recovery ${\cal R}_{{\rm A},t}$, is a linear function of the purities; therefore, from the error propagation, we have
\begin{align}
(\delta {\cal R}_{{\rm A},t})^2 =& \, \frac{1}{N_{\bo}} \EE_{\bo}[(P[\rho\sm(t)]-P[\rho\fil(t)]- {\cal R}_{{\rm A},t})^2] + \frac{1}{N_{\bo}}\EE_{\bo}[(\delta P[\rho\sm(t)])^2 ].
\end{align}
We note that, interestingly, this is not equal to a simple sum of the variance, i.e., $(\delta \bar{P}\sm(t))^2 + (\delta \bar{P}\fil(t))^2$, because the filtered and smoothed states (for each of the observed records) are strongly dependent on each other. For the latter quantity, the relative average purity recovery, ${\cal R}_{{\rm R},t}$, is a non-linear function of the three averages $\bar{P}\fil$, $\bar{P}\sm$, and $\bar{P}\god$, which are also not independent of each other. We thus need to calculate the variance from the full-form error propagation. Let us denote a quantity of interest by $f \equiv {\cal R}_{{\rm R},t} = (a-b)/(c-b)$, which is a function of $a$, $b$, and $c$, where each has its own variance denoted by $(\delta a)^2$, $(\delta b)^2$ and $(\delta c)^2$, respectively. The variance of ${\cal R}_{{\rm R},t} $ can be computed from
\begin{align}
(\delta {\cal R}_{{\rm R},t})^2 = (\delta f)^2 =&  \left( \frac{\partial f}{\partial a} \right)^2 (\delta a)^2 + \left( \frac{\partial f}{\partial b} \right)^2 (\delta b)^2 + \left( \frac{\partial f}{\partial c} \right)^2 (\delta c)^2 \nonumber + 2 \left( \frac{\partial f}{\partial a} \right) \left( \frac{\partial f}{\partial b} \right) \delta (a,b) \nonumber \\
& + 2 \left( \frac{\partial f}{\partial a} \right) \left( \frac{\partial f}{\partial c} \right) \delta (a,c) + 2 \left( \frac{\partial f}{\partial b} \right) \left( \frac{\partial f}{\partial c} \right) \delta (b,c),\\ 
= & \frac{1}{(c-b)^2} (\delta a)^2 + \frac{(a-c)^2}{(c-b)^4} (\delta b)^2 + \frac{(b-a)^2}{(c-b)^4} (\delta c)^2 \nonumber + 2 \frac{(a-c)}{(c-b)^3} \delta (a,b) \nonumber \\
& + 2 \frac{(b-a)}{(c-b)^3} \delta (a,c) + 2 \frac{(a-c)(b-a)}{(c-b)^4} \delta (b,c),
\end{align}
where
\begin{align}
a =&\,\, \bar{P}\sm(t) = \EE_{\bo}[P[\rho\sm(t)]],\\
b =&  \,\, \bar{P}\fil(t) = \EE_{\bo}[P[\rho\fil(t)]],\\
c = &\,\, \bar{P}\god(t) = \EE_{\bo}[P[\rho\god(t)]],\\
(\delta a)^2 = & \,\, (\delta {\bar P}\sm(t))^2, \\
(\delta b)^2 = & \,\, (\delta {\bar P}\fil(t))^2, \\
(\delta c)^2 = & \,\, (\delta {\bar P}\god(t))^2, \\
\delta(a,b) \equiv &\,\, \EE_{\bo}[ (P[\rho\sm(t)]- a)(P[\rho\fil(t)]- b)],
\end{align}
and in the similar way for $\delta(a,c)$ and $\delta(b,c)$. We note that $(\delta \bar{P}\god)^2$ is not zero for the cases of dOdU = dNdX and dYdX, where we have used $P(\rho\god) \rightarrow P(\rho\god^{\rm YZ})$ as discussed in the main text. Once we have the variances $(\delta {\cal R}_{{\rm A},t})^2$ and $(\delta {\cal R}_{{\rm R},t})^2$, we can compute the steady-state variances $(\delta {\cal R}_{\rm A,ss})^2$ and $(\delta {\cal R}_{\rm R,ss})^2$ using \eqref{eq-ss-err}.


\section*{References}

\bibliographystyle{iopart-num}

\providecommand{\newblock}{}

\end{document}